# Transport Properties of Topological Insulators Films and Nanowires*

Yi Liu(刘易)[a], Zheng Ma(马铮)[a], Yanfei Zhao(赵弇斐)[a], Meenakshi Singh[b], Jian Wang(王健)[a)b†]

[a]*International Center for Quantum Materials, School of Physics, Peking University, Beijing 100871, China*

[b]*The Center for Nanoscale Science and Department of Physics, The Pennsylvania State University, University Park, Pennsylvania 16802-6300, USA*

The last several years have witnessed the rapid developments in the study and understanding of topological insulators. In this review, after a brief summary of the history of topological insulators, we focus on the recent progress made in transport experiments on topological insulator films and nanowires. Some quantum phenomena, including the weak antilocalization, the Aharonov-Bohm effect and the Shubnikov–de Haas oscillations, observed in these nanostructures are described. In addition, the electronic transport evidence of the superconducting proximity effect as well as an anomalous resistance enhancement in topological insulator/superconductor hybrid structures is included.

**PACS**：73.20.Fz, 73.25.+I, 73.50.-h, 74.45.+c,

*Project supported by the National Basic Research Program of China (Grant Nos. 2013CB934600 & 2012CB921300), the National Natural Science Foundation of China (Grant Nos. 11222434 and 11174007), and the Pennsylvania State University Materials Research Science and Engineering Center under National Science Foundation (Grant No. DMR-0820404).

[†]Corresponding author. E-mail: jianwangphysics@pku.edu.cn

## 1  Introduction

Electronic transport measurements offer a unique insight into quantum mechanical phenomena in matter. For instance, the quantum Hall effect (QHE), in which the bulk was insulating and the edge was dissipationless, was first detected via electronic transport measurements in a two-dimensional electron gas confined in a strong magnetic field.[1]. This quantum Hall state is an example of a topological phase of matter, which, unlike the commonly known states of matter, is not characterized by a broken symmetry.[2,3]

A few decades after the discovery of the QHE, a new type of state, the quantum spin Hall (QSH) state, was theoretically predicted and experimentally observed in HgTe/CdTe quantum wells.[4-6] This novel state of matter, unlike QHE, does not require a magnetic field to manifest itself. The metallic edge states are mediated by a special band structure instead of a magnetic field. The concepts originally developed for the QSH state in HgTe/CdTe quantum wells were soon generalized to three dimensional (3D) topological insulators (TIs), which are conducting on the surface and insulating in the bulk. The $Bi_{1-x}Sb_x$ alloy was the first predicted 3D TI.[7] The band structure of TIs has a band gap in the bulk, but no such gap on the surface. The surface conduction states, discussed in more detail below, form Dirac cones. The evidence of these Dirac cones was forthcoming with the development of surface probe techniques, such as the angle-resolved photoemission spectroscopy (ARPES). An odd number of Dirac cones in the surface conducting states along with a bulk band gap were observed in ARPES measurements on a $Bi_{1-x}Sb_x$ alloy proving it to be the much sought after TI.[8] However, the complicated band structure of that material hindered further experimental studies. Fortunately, 3D TIs with simpler band structures were predicted in $Bi_2Te_3$, $Bi_2Se_3$, and $Sb_2Te_3$ through first-principle calculations[9] and were confirmed by ARPES measurements.[10,11] The ARPES measurements, in agreement with the theoretical predictions, showed a continuous energy spectrum of conducting surface states with a linear dispersion relation and a single Dirac cone. Because of their large bulk gaps and simple band structures, these "new generation" TIs have been drawing more and more attention and ingenious theoretical and experimental studies are contributing exciting in this rapidly developing field.[12,13]

Like ordinary insulators, 3D TIs have a band gap between conduction and valence bands in the bulk. However, unlike ordinary insulators, they possess a gapless metallic surface state. This exotic surface state is protected by the time-reversal invariant symmetry (TRIS). In layman terms, this means that if the progression of time is reversed, the new state cannot be told apart from the old. As a consequence of this TRIS, the momentum and spin of a surface state are locked. This means that electrons of a certain spin can only travel in a certain direction and back-scattering is prohibited. In band structure terminology, each momentum along the surface has only a single spin state at the Fermi level. This unusual property of TIs was confirmed by spin-ARPES experiments.[14,15] The surface states protected by TRIS are robust to disorder and non-ferromagnetic impurities, which shows the wide potential for proposed applications in spintronics devices and other fields.

It is noteworthy that most of the historical developments cited so far involve the ARPES technique. The reason for this is that a surface sensitive technique like the ARPES is ideal for probing interesting surface properties. Transport measurements on the other hand, tend to probe the bulk more than the surface simply because the bulk has more electrons to contribute to the transport. At first, this may not seem like a problem since the bulk of a TI has a band gap. The Fermi level, however, may not lie in this gap in a real experimental situation leading to non-negligible bulk conduction. This makes experimental detection of the surface states by transport measurements much more challenging. Nevertheless, considerable efforts – mainly involving the control of the Fermi level by doping and electrical gating, have been made to overcome these difficulties. For example, in $Bi_2Se_3$, which is a large band gap (0.3 eV) TI, the Fermi level lies in the bulk conduction band due to the electron-type bulk carriers induced by Se vacancies. This situation can be rectified by doping and electrical gating, tuning the Fermi level into the bulk band gap and largely suppressing the bulk conductance.[16] Alternatively, some quantum phenomena in electronic transport may offer signature dependence on only surface related properties and be used as the evidence of surface dominated transport. For example, the periods of Shubnikov–de Haas (SdH) oscillations have been found to depend only on $H_\perp$ (the magnetic field perpendicular to the surface), which indicates that the transport is supported by the 2D surface states. [17]

In this review article, we focus on transport experiments performed by our group on TI thin films and nanowires as well as TI/superconductor (SC) hybrid structures. In Section 2, experiments on TI thin films are described. A combination of weak antilocalization (WAL) and electron-electron interaction (EEI) is used to explain the experimental results. Anomalous anisotropic mangetoresistance in TI thin films is also reviewed in this Section. Section 3 summarizes quantum oscillations in TI nanoribbons and nanowires, including SdH oscillations and Aharonov–Bohm(AB) oscillations. Section 4 is devoted to TI/SC hybrid structures, which are a platform to detect Majorana Fermions.[18] The evidence of superconducting proximity effect in TI nanoribbons and anomalous resistance enhancement in TI/SC hybrid film structures are shown. A brief conclusion and outlook are given in Section 5.

## 2  Transport properties of TI thin films

Progress in thin film growth techniques, such as molecular beam epitaxy (MBE), makes the layer by layer growth of high quality TI thin films possible.[19-21] There is a lower bound to the thickness of a 3D TI film, below which the coupling between the top and the bottom surface occurs. This bound is material dependent and varies from 6 quintuple layers (QL, 1QL is around 1 nm thick) in $Bi_2Se_3$ to 4QL in $Sb_2Te_3$ and 2QL in $Bi_2Te_3$.[22-25] TI thin films are amenable to doping and gating, which can be used to ensure that the Fermi level lies in the bulk band gap and thereby reduce the bulk conductance. The surface states in thin films are also supposed to dominate the transport by virtue of the films' high surface-to-volume ratios. Thus, TI films are the preferred platform for many theoretical and experimental studies.[26-30]

### 2.1  Weak antilocalization and electron- electron interaction

Electrons in a TI surface state acquire a Berry phase of $\pi$ after traversing a closed trajectory above the Dirac cone. The destructive interference between a pair of closed and time-reversed paths, resulting from the $\pi$ Berry phase, leads to a conductivity maximum in zero magnetic field, which underlies the WAL. The effect is suppressed by applying a magnetic field, thus giving rise to negative magnetoconductivity. Because of the robustness of the surface state, the WAL is protected against the strength of disorder and nonmagnetic impurity. Doping with magnetic impurities, however, breaks the time-reversal symmetry and destroys the TI state. Such doping therefore opens an energy gap in the surface states, causing a crossover from the WAL to regular weak localization (WL).[31] The WAL in TI thin films was first revealed in transport experiments.[32-35] Among these studies, the work of our group on pure and doped $Bi_2Se_3$ thin films, for the first time, quantitatively fitted the magnetoresistance and resistance vs. temperature results in low temperature regime by combining the EEI with the WAL effect in TI films.[32] The 45-QL $Bi_2Se_3$ thin films were grown by MBE on insulating 6H-SiC (0001) substrates. According to the ARPES results, the Dirac point of the pure $Bi_2Se_3$ film is located at 0.185 eV below the Fermi level, which means that both the surface states and the bulk conduction band contribute to the electronic transport. The Fermi level of our doped $Bi_{2-x}Pb_xSe_3$ thin film, however, was tuned into the bulk band gap by doping (see Fig. 1).

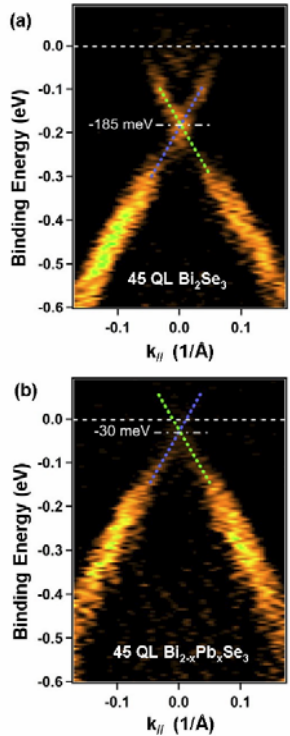

**Fig. 1.** ARPES spectra of the 45-nm thick crystalline films: (a) undoped $Bi_2Se_3$ film, (b) doped $Bi_{2-x}Pb_xSe_3$ film. The energies are measured relative to the Fermi level. [32]

The temperature dependences of the conductance (G) for the $Bi_2Se_3$ and $Bi_{2-x}Pb_xSe_3$ samples are shown in Figs. 2(a) and 2(b), respectively. The logarithmic temperature dependence of the conductance can be fitted by a 2D WL theory. The WL theory however, predicts a positive magnetoconductivity in a perpendicular field.[36] The Comparison of the experimentally measured conductance ($\Delta G = \Delta G(H) - \Delta G(0)$) as a function of magnetic field and the theoretical expectation from the WL theory is shown in Figs. 2(c) and 2(d). It is evident that experiment and theory do not agree. Figures 2(e) and 2(f) show the magnetoconductance in a parallel field along with the theoretical fit.[37]

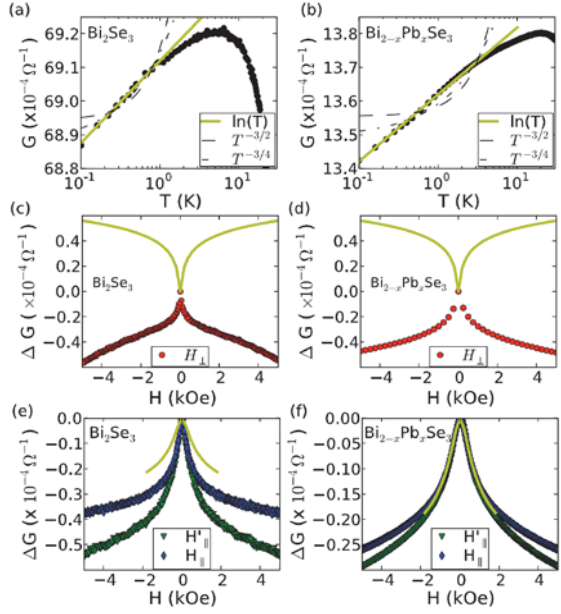

**Fig. 2.** WL theory. (a) and (b) Temperature dependence of the conductance. Lines are fits to the WL theory in 2D (solid line) and 3D (dashed and dash-dotted lines, corresponding to phonon and electron-electron scattering as the source of dephasing). The temperature dependence of both films suggests that our films are in the regime of weak spin-orbit scattering. However, in this regime, the theory predicts a positive magnetoconductance in perpendicular field. (c) and (d) Change in conductance, $\Delta G = \Delta G(H) - \Delta G(0)$. The theory predicts a localization effect; however, we observe antilocalization. (e) and (f) Magnetoconductance in parallel fields, $H_{\parallel}$ denotes a field parallel to the film and perpendicular to the excitation current, while $H_{\parallel}'$ denotes a field parallel to both the film and the current. Lines are best fits to the theory.[32]

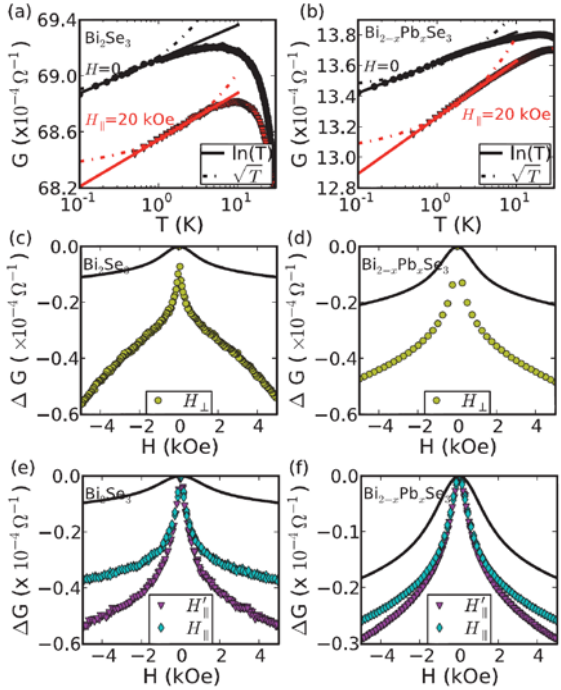

**Fig. 3.** The electron-electron interaction. (a) and (b) Temperature dependence of the conductance. Lines are fits to the 2D (solid line) and 3D (dashed line) theory. Panels (c), (d) and (e), (f) show magnetoconductance in perpendicular and parallel magnetic fields, respectively. Solid lines in panels (c)-(f) are fits to the EEI theory using no additional fitting parameters besides those found in panels (a) and (b).[32]

The contradiction between WL theory and the experimental data urged us to take EEI into account. The experimentally measured temperature and magnetic field dependences of the conductance along with theoretical fits are shown in Fig. 3. The theoretical fits with the EEI correctly reproduce the signs of magnetoconductance in the perpendicular field. This shows the importance of the EEI in TI thin films. However, the theoretical fits still do not exactly match the magnetoconductance results both in perpendicular and parallel fields (see Figs. 3(c) - (f)). Therefore, we combined WAL with EEI to fit our experimental results. In Fig. 4, we can see that the experimental data of both *G-T* and *ΔG-H* behaviors are fitted well to the combined model in the low temperature and small field regime. The revealed WAL and EEI effect in $Bi_2Se_3$ thin films has been further confirmed by many other experiments in TI thin films and nanoribbons.[33,35,38,39]

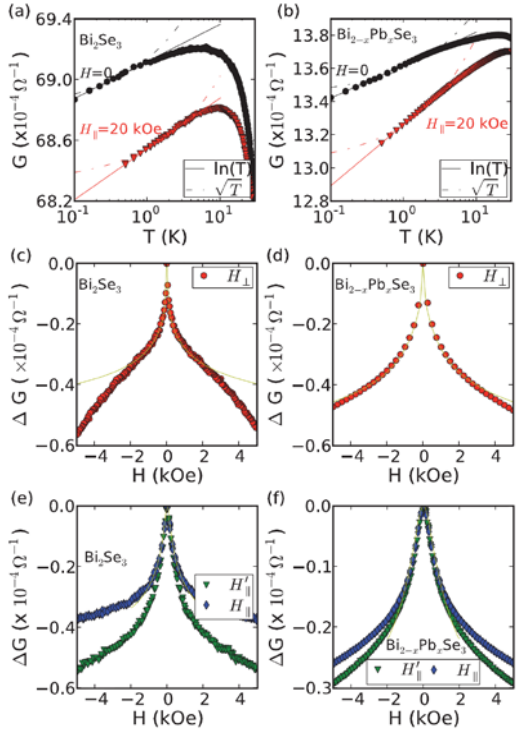

**Fig. 4.** Transport properties in low temperature and low magnetic field. The left column is for the undoped $Bi_2Se_3$ film, while the doped $Bi_{2-x}Pb_xSe_3$ data are shown on the right column. (a) and (b) Conductance versus temperature for zero magnetic field (black circles) and $H_\parallel$ = 20 kOe (red triangles), with fits to 2D theories (solid lines) and 3D EEI theory (dash-dotted lines). (c) and (d) Magnetoconductance $\Delta G = \Delta G(H) - \Delta G(0)$ for fields perpendicular to the film. Solid lines are the result of a combined WAL and EEI, as explained in the text. (e) and (f) Magnetoconductances for in-plane fields $H_\parallel^{'}$ and $H_\parallel$, which are parallel and perpendicular to the current

direction, respectively. The solid lines are fits to the combined WAL and EEI model. The magnetoconductance data are all taken at 500 mK.[32]

### 2.2 Anisotropic magnetoresistance

The measurement of magnetoresistance, is a direct and key avenue to study quantum phenomenon like the WAL described above. In a separate experiment on TI thin films,[40] a small MR dip (WAL) was also observed at low fields, while an anomalous anisotropic MR was found at higher fields, different from all the previously observed behaviors. More than 10 samples were measured in order to confirm and study this phenomenon. Four of them, marked sample 1-4, will be discussed here. Samples 1, 2 and 3, grown on high resistivity silicon substrates by MBE, are 200-QL single crystal $Bi_2Se_3$ films. Among them, samples 2 and 3 came from the same film with Hall bars oriented along and perpendicular to the crystal axis [-110], respectively. Sample 4 is a 45-QL $Bi_2Se_3$ film grown by MBE on a sapphire substrate and covered by 20-nm thick Se protection layer.

The normalized MR of sample 1 under different in-plane fields is shown in Fig. 5(a) and 5(c). Figures 5(b) and 5(d) show the temperature dependence of these MRs. The anisotropy in field is apparent. When the field is perpendicular to the current direction and crystal axis [110], the MR is positive and about 10 times smaller than the MR when the field is perpendicular to the film. When the field is aligned along the excitation current as well as the crystal axis [110], the MR is negative other than the MR dip arising from WAL near the zero field.

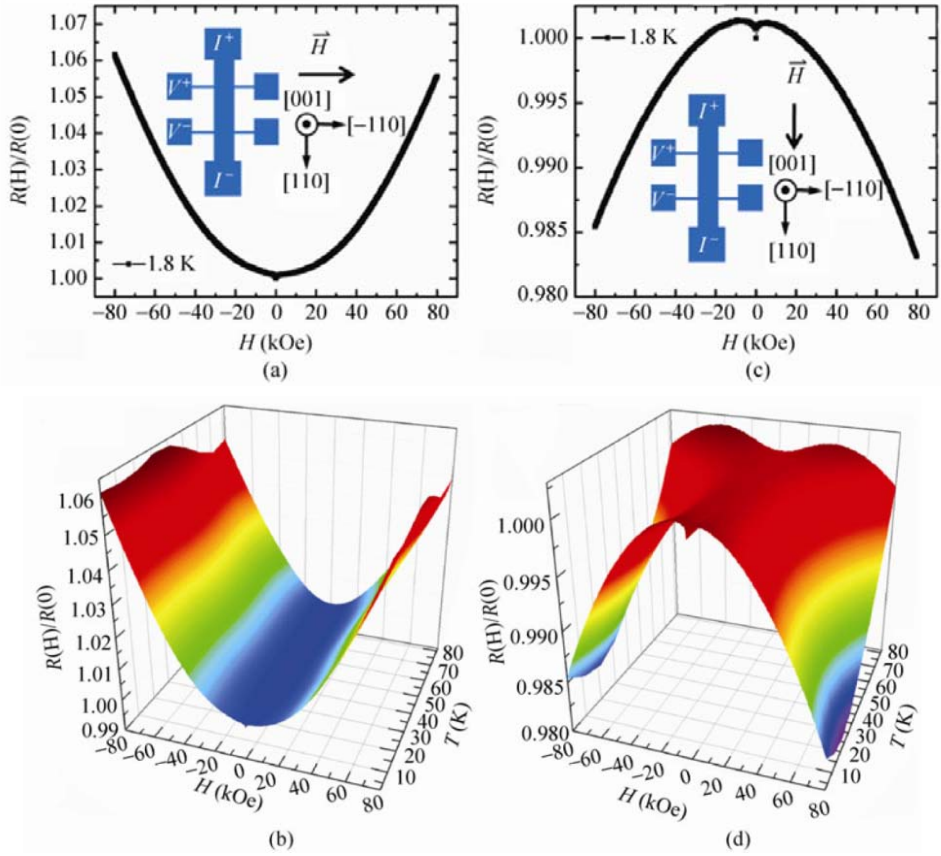

**Fig. 5.** In-plane magnetoresistance of sample 1 (200-nm thick $Bi_2Se_3$ film). (a) The magnetic field is perpendicular to the current direction and crystal axis [110] of sample 1 at 1.8 K. (b) 3D image of the magnetoresistance at different temperatures when the field is perpendicular to the current. (c) The magnetic field is parallel to the current direction and crystal axis [110]. (d) 3D image of the magnetoresistance at different temperatures when the field is parallel to the current.[40]

From the above results, it is unclear whether the anisotropic MR depends on the direction of the crystal axis. A control experiment was therefore performed on samples 2 and 3. Due to the special orientation of the Hall bar structures, the excitation current propagates along crystal axis [-110] in sample 2 and perpendicular to [-110] in sample 3. Irrespective of the crystal axis direction, the MR of the $Bi_2Se_3$ film was positive when the magnetic field was perpendicular to the current and negative when the field was parallel to the current, as shown in Fig. 6. Further measurements on sample 4 (a 45-QL $Bi_2Se_3$ film) confirmed this phenomenon (See Fig. 7). The temperature dependence of the anisotropic MR behavior is also shown in Fig. 7. Increasing temperature with the magnetic field parallel to the current lowers the positive MR and increases the threshold magnetic field of the negative MR situation, in agreement with the result for sample 1 (Fig. 5(d)).

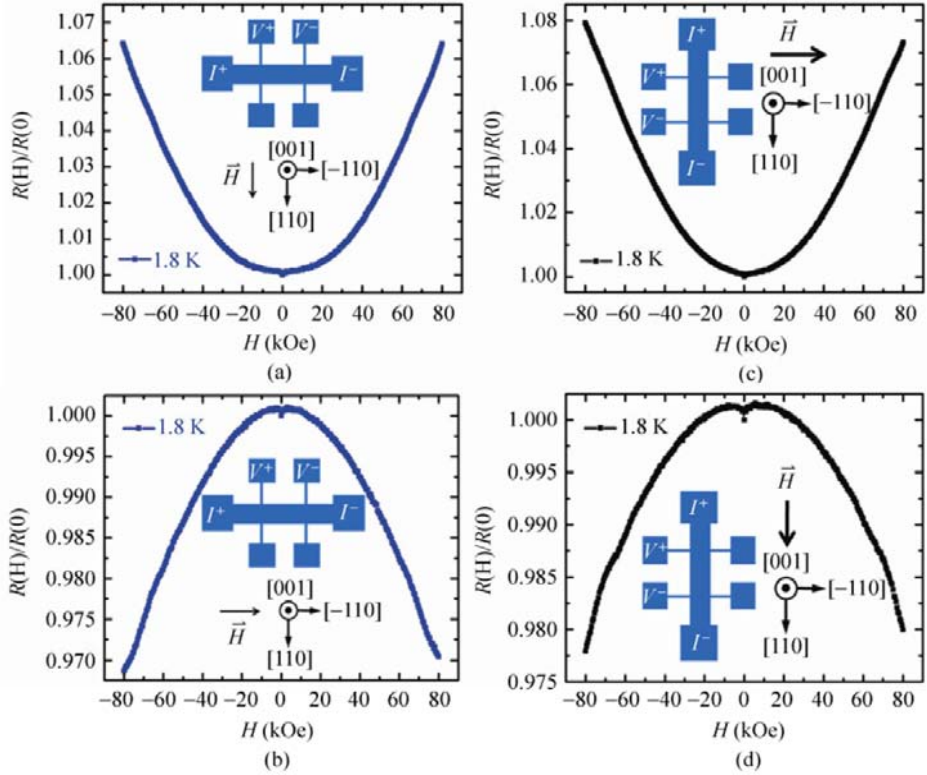

**Fig. 6.** In-plane magnetoresistance of samples 2 and 3 (the two samples are from the same 200-nm thick $Bi_2Se_3$ film). (a) The magnetic field is perpendicular to the current and crystal axis [-110] of sample 2 at 1.8 K. (b) The field is parallel to the current and crystal axis [-110] of sample 2. (c) The magnetic field is perpendicular to the current direction and crystal axis [110] of sample 3. (d) The magnetic field is parallel to the current direction and crystal axis [110] of sample 3.[40]

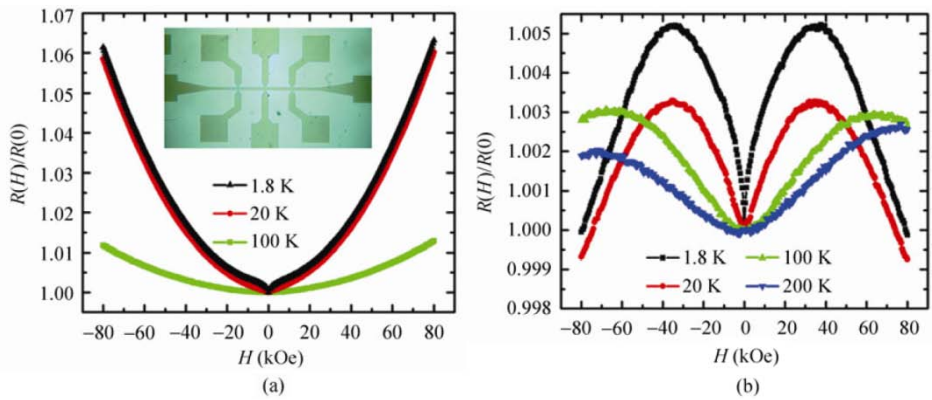

**Fig. 7.** (a) In-plane magnetoresistance at different temperatures when the field is perpendicular to the current. The inset is an optical image of the Hall bar structure used. The width of the Hall strip is 40 μm and the distance between two adjacent Hall bars is 400 μm. (b) In-plane magnetoresistance at different temperatures when the field is parallel to the current.[40]

Although a theoretical explanation of the anisotropic MR still remains elusive, we have ruled out some possible explanations by contrasting our results with some typical anisotropic MR experiments on narrow band gap semiconductors, such as a detailed study on InSb thin films.[41] This indicates that neither a classical reason (skipping orbits) nor a quantum explanation (WL) can perfectly interpret our transport experiment and an explanation involving the special spin-helical surface state of TIs may be in order. A separate study[42] has also reported a negative MR in Sn-doped $Bi_2Te_3$ films in parallel field. However, the orientation of the magnetic field relative to the applied current for the negative MR is opposite to our experiment.

More efforts are necessary to acquire a thorough understanding of the results presented above, including figuring out the collective effect of the surface and bulk states, higher-order corrections for the surface states, as well as the orbital effects. Measuring angular dependence of the MR and studying electrically-gated samples might be helpful.

## 3  Quantum oscillations in TI nanoribbons and nanowires

TI nanoribbons and nanowires are ideal candidates to study transport properties of the surface states due to their large surface to volume ratio. In mesoscopic structures, the wavefunctions of electrons remain coherent when propagating over relatively long distances at low temperatures. If we consider two interfering partial electron waves enclosing a magnetic flux $\Phi$, the phase of the two electron waves will be changed by the magnetic flux, thus turning a constructive interference into a destructive one, and vice versa. According to a seminal work by Aharonov and Bohm in 1959,[43] the phase difference between the two waves is $4\pi\Phi/\Phi_0$, where $\Phi_0 = h/e$ is the magnetic flux quanta. It is the so called Aharonov-Bohm effect (AB effect), which is accompanied by observable periodic MR oscillations.[44] The AB effect turns out to be a sensitive tool for probing the surface transport in TI nanoribbons with a magnetic field applied along the ribbon. If the electron dephasing length exceeds the sample diameters, quantum interference will arise in the electron trajectories. For electrons traveling through the bulk of the ribbon, the interference loops depend on the positions of impurities and the enclosed magnetic fluxes varied, ruling out the periodic MR oscillations. For electrons traveling along the TI surface, on the other hand, each coherent trajectory taking part in the quantum interference encloses the same area perpendicular to the magnetic field. Therefore, a periodic MR oscillation, which is the hallmark of AB effect, occurs in TI nanoribbons with the characteristic period of the external magnetic field $\Delta H = \Phi_0/S$, where $\Phi_0 = h/e$ is the flux quantum and S is the cross-sectional area of the nanoribbon. It has

been theoretically and experimentally studied by several groups.[44-46]

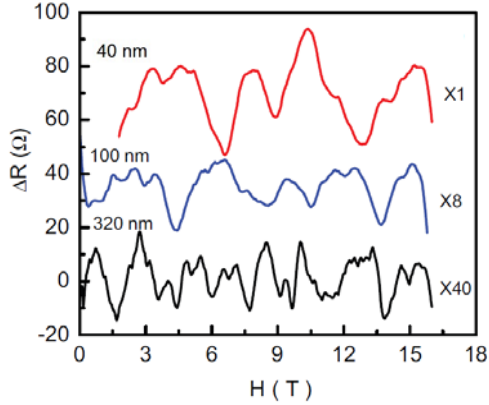

**Fig. 8.** MR variations, $\Delta R$ vs. $H_\perp$ curves by subtraction of a smooth background in 40-nm, 100-nm and 300-nm diameter nanowires at 2.5 K, 2K, and 2 K respectively. The amplitudes of the fluctuations or quantum oscillations increase with the decrease of wire diameter.[47]

Recently, the transport experiment of cylindrical single-crystal $Bi_2Te_3$ nanowires was carried out, which revealed the dual evidence, including AB effect and SdH oscillations, for the Dirac surface state in TI nanowires.[47] A systematic comparative study on $Bi_2Te_3$ nanowires of various diameters was carried out in that work. Nanowires of three different diameters (40 nm, 100 nm and 300 nm) were fabricated using template-assisted electrodeposition. The MR of these nanowires was measured in a perpendicular magnetic field, and the amplitudes of the MR fluctuations or oscillations in 40-nm wire at 2.5 K were much larger than those in 100-nm and 300-nm wire at 2K, as shown in Fig. 8. To gain further insight to this phenomenon, we mainly focus on the data of 40-nm wire.

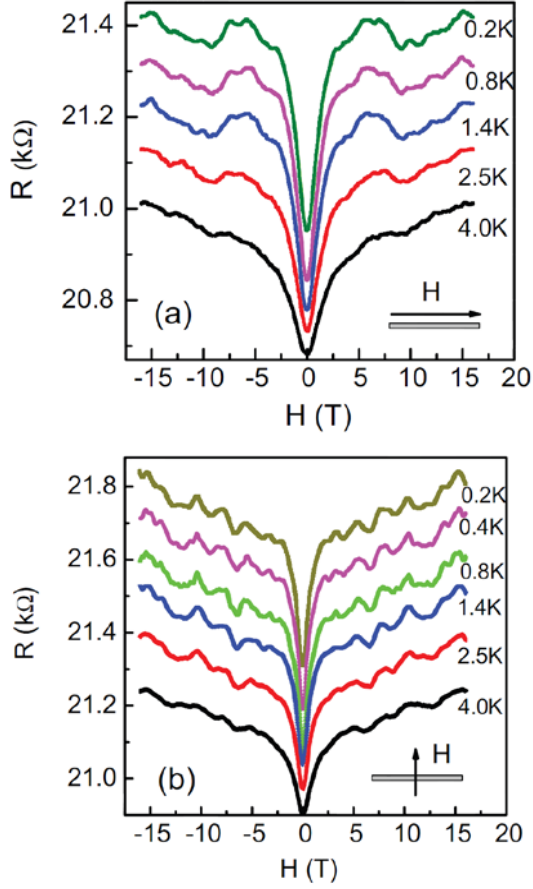

**Fig. 9.** The (a) $R$ versus $H_\parallel$ and (b) $R$ versus $H_\perp$ curves of the 40-nm wire measured at different temperatures.[47]

In Fig. 9, MR oscillations of the 40-nm wire can be observed in both perpendicular and parallel magnetic field at various temperatures. After the subtraction of a background by fitting the original data with a polynomial function, we find that the MR oscillations in parallel magnetic field are periodic with $\Delta H_\parallel = 4.45$T. (See Fig. 10(a)). According to the formula $\Delta H_\parallel = \Phi_0/S$, we can evaluate the diameter of the nanowire, which is 35nm. Considering the oxide layer on the wire surface, the result is consistent with the actual wire diameter. The experimental data provide evidence that the MR oscillations in a parallel field in the $Bi_2Te_3$ nanowire, known as the AB effect, come from the metallic Dirac surface states. Besides the AB effect, some tiny dips are also observed in Fig. 10, which are found to superimpose on the peak of AB oscillations. One possible explanation of these tiny dips is the normal Altshuler-Aronov-Spivak (AAS) effect with $\Phi_0/2$ period, half of the period of AB effect. These dips are relatively weak and occur only at odd multiples of $\Phi_0/2$ since the dips at even multiples of $\Phi_0/2$ are exactly at the same positions of the AB effect, thus cannot be observed. For the topological nontrivial surface state, the AAS effect is suppressed due to the spin helical structure. Our observation of the AB oscillations, combined with the relatively weak AAS oscillations, supports the existence of the topological nontrivial Dirac state on the surface of $Bi_2Te_3$ nanowires.

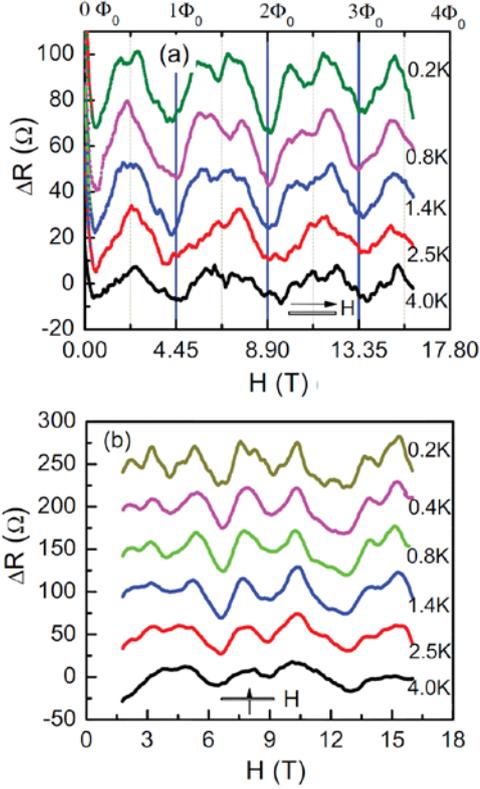

**Fig. 10.** Panel (a) shows the variation of the resistance, $\Delta R$, versus $H_\parallel$ measured at different $T$ after the subtraction of a background by fitting the original data with a polynomial function. Panel (b) shows the variations of the resistance, $\Delta R$, versus $H_\perp$ after the subtraction of a background measured at different $T$.[47]

The MR oscillations in the perpendicular field show another evidence of the TI surface state.[47] When a material is subjected to a magnetic field, continuous energy bands are quantized to Landau levels (LLs), and quantum oscillations appear in the MR of the material in period of $1/H$. This quantum phenomenon, named SdH oscillation, was firstly observed in single crystal Bi in 1930.[39] The SdH oscillations originate from the successive emptying of the LLs as the magnetic field is increased. Nowadays, several groups have succeeded in revealing the SdH oscillations in the 2D surface state of TIs.[18, 48, 49]

The Landau index n is relative to the extremal cross section of Fermi surface $S_F$, described by $2\pi(n + \gamma) = S_F \hbar/eH_n$,[18] where $H_n$ is the magnetic field when the n-th LL is filled by the electron. Parameter $\gamma = 0$ or $-1/2$, indicating two typical cases. $\gamma = 0$ is for the Schrodinger spectrum, where there are nfilled LLs below the Fermi surface; $\gamma = -1/2$ is for the Dirac case, where there are n+1/2 filled LLs between the Fermi energy and the Dirac point. The conduction and valence band both contribute half of the 0-th LL, leading to the 1/2-shifted SdH oscillation.

In Fig. 10(b), we find that the MR oscillations are not periodic with perpendicular magnetic field $H_\perp$, and some small kinks appear at 0.2K. Then we plot the variations of the resistance, $\Delta R$ versus $1/H_\perp$. If we consider the dips and small kinks at 0.2 K, a periodic oscillation with the period of $\Delta(1/H_\perp) = 0.036 \text{T}^{-1}$ becomes apparent, as indicated by red dashed lines in Fig. 11(a). Figure 11(b) plots the linear relation between Landau index n and the inverse of the perpendicular magnetic field $1/H_\perp$. The intercept of curve A is near -0.5, which indicates a 1/2-shifted SdH oscillation and provides another transport evidence for TI surface Dirac spectrum in $Bi_2Te_3$ nanowires.

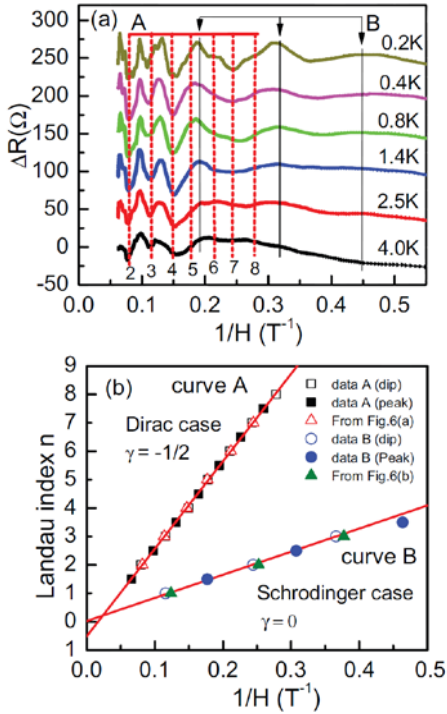

**Fig. 11.** Panel (a) shows the variations of the resistance, $\Delta R$, versus $1/H_\perp$ after the subtraction of a fitted smooth background at different *T*. In panel (b), curve A plots the integer LL index, n (open squares), versus $1/H_\perp$ at the minima or kinks of the SdH spectrum. The data defined from the maxima are also shown additionally in the plot as the solid squares (shifted by 1/2 in LL index). Curve B plots the integer LL index, *n* (open squares) versus $1/H_\perp$ for another set of SdH spectra in low-H regime.[47]

To summarize, the observation of the AB oscillations and the 1/2-shifted SdH oscillation provide the dual evidence of topological nontrivial surface states in $Bi_2Te_3$ nanowires, which is a first experimental demonstration of TI surface state in cylindrical nanowires.

## 4 TI and SC hybrid structures

Unlike the newly discovered TIs, superconductivity has remained one of the most important research areas in condensed matter physics and material science since its discovery more than 100 years ago. This is not only because the zero-resistance of SCs showed great opportunity to be used as electric cables (e. g. coils in superconducting magnet), but also because hybrid structures of SCs with other materials are powerful tools for exploring the fundamental physics as well as innovative applications. The most significant example of SC hybrid structure is the Josephson junction. [50] In the Josephson Effect, when two SCs are separated by a constriction, a supercurrent can flow through the constriction without any applied voltage. The Josephson Effect is the foundation for the superconducting quantum interference devices (SQUID) whichare powerful tools to measure magnetic fields and are widely used in scientific researches, medicine and industry.

The combination of SCs and TIs also holds several promises for fundamental physics and applications. [51] The long sought Majorana fermions for example, are predicted to exist in the interface of SC/TI hybrid structures. [18] The Majorana fermion is its own antiparticle and can be used in error-tolerant topological quantum computation. Fu and Kane [52] have pointed out that when an s-wave SC combines with a TI, Majorana fermions can emerge in the superconducting proximity induced vortices. The Proximity effect at TI and SCjunctions was first observed in the interface between a SC and $Bi_{1-x}Sb_x$ [53] using transport measurements. The material was proved to be the first generation of 3D TIs later. Recently, the proximity induced superconducting gap in the TI surface was also observed by STM. [54] The sample geometry in that experiment also makes the direct observation of Majorana-bound states possible.

### 4.1 Superconducting proximity effect in TIs

In the superconducting proximity effect, when a SC is in contact with a non-SC, the wave function of superconducting electrons 'leaks' through the interface making the non-superconductor part near the interface superconducting. A systematic study of proximity effect in TI nanoribbons by transport measurements has been carried out by our group. [38]

We measured the transport properties of mesoscopic $Bi_2Se_3$ nanoribbons contacted by superconducting tungsten leads. It was found that the proximity effect induced superconducting length scale in $Bi_2Se_3$ was much longer than that predicted by diffusion channel model. Such a long range proximity effect cannot be realized in the diffusive bulk channel of $Bi_2Se_3$ nanoribbons. If we consider the ballistic surface state channels in TI nanoribbons however, the experimental results become reasonable. Furthermore, the residual resistance of the nanoribbon below the superconducting transitition temperature shows periodic oscillations with the magnetic field, which is hard to explain in the traditional framework.

Our $Bi_2Se_3$ nanoribbons were made based on the gold catalyzed vapor-liquid-solid mechanism by using a horizontal tube furnace.[44] The electron diffraction pattern and Raman peak have shown that the growth direction is along [11$\bar{2}$0] and the samples are single crystals with little disorder. We deposited both superconducting tungsten (W) and normal platinum (Pt) electrodes using a dual beam focused ion beam (FIB) etching and deposition system. The W electrodes, which are composed of tungsten, carbon and gallium, show $T_c$ around 4.7 K (much larger than 15 mK of pure W) and a large critical field ($H_c$~70 kOe)[55], which is beneficial for the vortex study. We have made several devices in the way mentioned above, and their parameters are shown in Table 1.

TABLE I. Summary of $Bi_2Se_3$ nanoribbon devices.

| Device | Length ($\mu$m) | Width (nm) | Thickness (nm) | Contact width (nm) | Resistivity at 6 K (m$\Omega$ cm) | Contact material |
|---|---|---|---|---|---|---|
| A | 1.08 | 600 | 60 | 200 | 0.51 | W |
| B1 | 0.94 | 430 | 60 | 430 | 1.37 | W |
| B2 | 1.55 | 430 | 60 | 430 | 1.37 | W |
| C | 2.47 | 284 | 276 | 316 | 0.79 | W |
| D | 2.30 | 240 | 77 | 240 | 0.70 | W |
| E | 5.12 | 300 | 98 | 122 | 0.69 | W |
| F | 1.00 | 700 | 60 | 300 | 3.53 | W |
| G | 5.73 | 270 | 60 | 234 | 0.39 | Pt |
| H | 0.94 | 370 | 50 | 280 | 0.48 | Pt |

First, we studied two devices with Pt electrodes in the four-probe geometry. The relationship between conductivity and temperature is shown in Fig. 12(a) and Fig. 12(b). The logarithmic temperature dependence of conductivity is due to WAL and EEI. The MR shown in Fig. 12(c) and Fig. 12(d) can also be understood in this framework. These curves can be fitted by the

Hikami-Larkin-Nagaoka equation [36] in which the value of parameter α can tell how strong the spin-orbit coupling is. In our case, α is around -1.3, impling that there is more than one parallel conducting channel contributing to the conductivity.

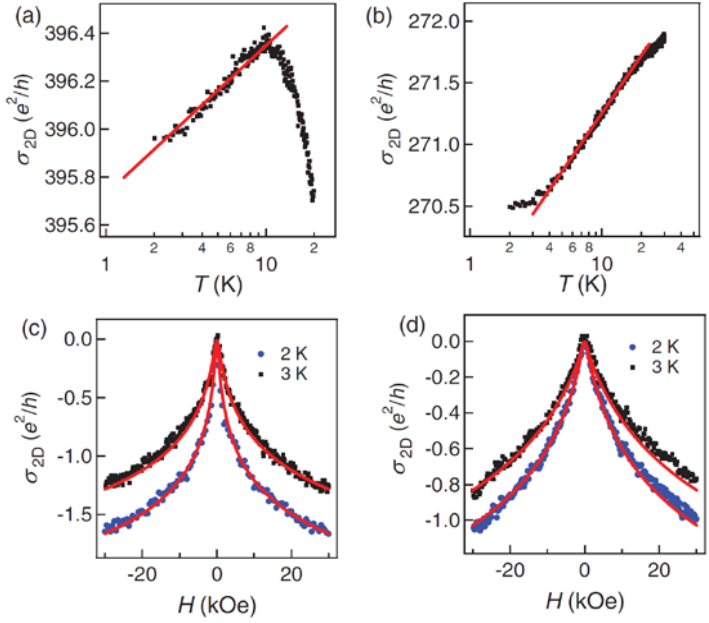

**Fig. 12.** (Color online) (a) Temperature-dependent conductivity for device G. The solid line is a ln( T ) fit. (b) Temperature-dependent conductivity for device H. The solid line is a ln( T ) fit. (c) Magnetoconductivity for device G at T = 3.0 K (squares) and T = 2.0K (circles). The solid lines are fits to the Hikami-Larkin-Nagaoka theory with EEI included. (d) Magnetoconductivity for device H at T = 3.0 K (squares) and T = 2.0 K (circles). The solid lines are fits to the Hikami-Larkin-Nagaoka theory with EEI included.[38]

Then a nanowire device (device A, inset of Fig. 13(a)) with superconducting W electrodes was studied. The temperature dependence of the resistance was measured (Fig. 13(a)). The result clearly showed the proximity effect when the W electrodes became superconducting at around 4.7 K. When the temperature dropped below 2 K, the device attained zero resistance. The V vs. I behavior is shown in Fig. 13 (b). The "subharmonic gap structure" of dI/dV at 500 mK shown in Fig. 13(c) is due to multiple Andreev reflections at $V = \frac{2\Delta}{ne}$ (1), where $\Delta$ is the superconducting gap of the W electrodes and n is an integer. The Andreev reflections in S-TI-S junctions were also observed in another transport experiment.[56] From Eq. (1) we can deduce that $T_c$ = 4.41 K, which is consistent with the $T_c$ of the W electrodes.

Normally, in order to observe the supercurrent and multiple Andreev reflections in a SC-Normal metal-SC (SNS) device, the normal channel length L must be shorter than the thermal length $\xi_N$ and electron phase-breaking length $L_\phi$. In our experiment, the $\xi_N$ calculated by using carrier density in $Bi_2Se_3$ and diffusion channel model (70 nm) is much smaller than the $Bi_2Se_3$ channel length (around 580 nm) in our W-$Bi_2Se_3$-W hybrid structure. Thus, the multiple Andreev reflections should not be observed in our samples. However, if we consider the coexistence of ballistic (surface) and diffusive (bulk) transport channels in $Bi_2Se_3$ nanoribbons, the multiple Andreev reflections become possible since $\xi_N$ is near 340 nm in ballistic limit at 1.8K, which is comparable to the channel length.

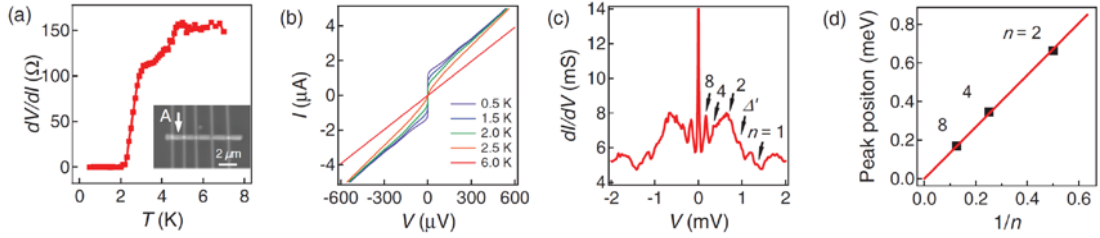

**Fig. 13.** (Color online) (a) Two-probe zero-bias dV/dI vs temperature for device A at H = 0. Inset shows an SEM image of the device, the arrow indicats the measured channel with edge-to-edge length of 1.08 μm between two W electrodes. (b) The I -V characteristics of device A at various temperatures, measured using the same contacts as in (a). (c) dI/dV vs. V in device A at T = 500 mK and in zero magnetic field. The arrows identify a consistent subharmonic series of conductance anomalies corresponding to subharmonic gap structure (n = 2, 4, 8). (d) Position of differential conductance anomalies as a function of the index 1/n. [38]

Another noteworthy phenomenon is the zero-bias conductance peak (ZBCP) (see Fig. 13(c)). The ZBCP has also been observed in $Bi_2Se_3$ films with gold electrodes and doped TI $Cu_xBi_2Se_3$.[57][58] This ZBCP can be explained in the framework of topological superconductors (TSCs) which are characterized by the presence of Majorana fermions. In the theory of TSC, [59] the appearance of ZBCP in conductance is due to Majorana fermions and reflects their zero-energy bound state. However, the observed ZBCP is not a stand-alone demonstration of the existence of Majorana fermions in these systems.

Novel MR oscillations were observed in our samples when the W electrodes became superconducting. The magnetic field was applied perpendicular to the nanoribbon and the substrate. As shown in Fig. 14(a) and Fig. 14(b), the B1 channel (inset of Fig. 14(a)) of the $Bi_2Se_3$ device shows periodic MR oscillations excluding the low field data. Figure 14(b) shows that the oscillation amplitude varies non-monotonically with the temperature and disappears in both high and low temperature regions. Figures 14(c) and 14(d) present the measurement results obtained in the B2 channel (see inset of Fig. 14(a)), the oscillation pattern is the same as that of the B1 channel. In other control measurements, we found that the periodic oscillations only appear when the magnetic field is perpendicular to the nanoribbon.. This behavior accompanied with the nonmonotonic temperature dependence of the MR oscillations rules out the possibility of the AB effect. The SdH effect can also be excluded since the period of SdH oscillation is in 1/H not H. After ruling out several possible mechanisms, we finally turn to the "Weber blockade" model.[60] It is a theory developed for pure SCs, but we assume it can also be applied in this case. In this model, the perpendicular magnetic field controls the number of Pearl vortices in the superconducting strip, corresponding to proximity induced superconducting $Bi_2Se_3$ in our case. As shown in Fig. 14(e) and (f), the theoretical calculation agrees well with our experimental results. Thus, in our experiments we observed the long range proximity effect in the TI surface and the possible induced vortices in the interface between TI and SC, which paves a way to pursue Majorana fermions.

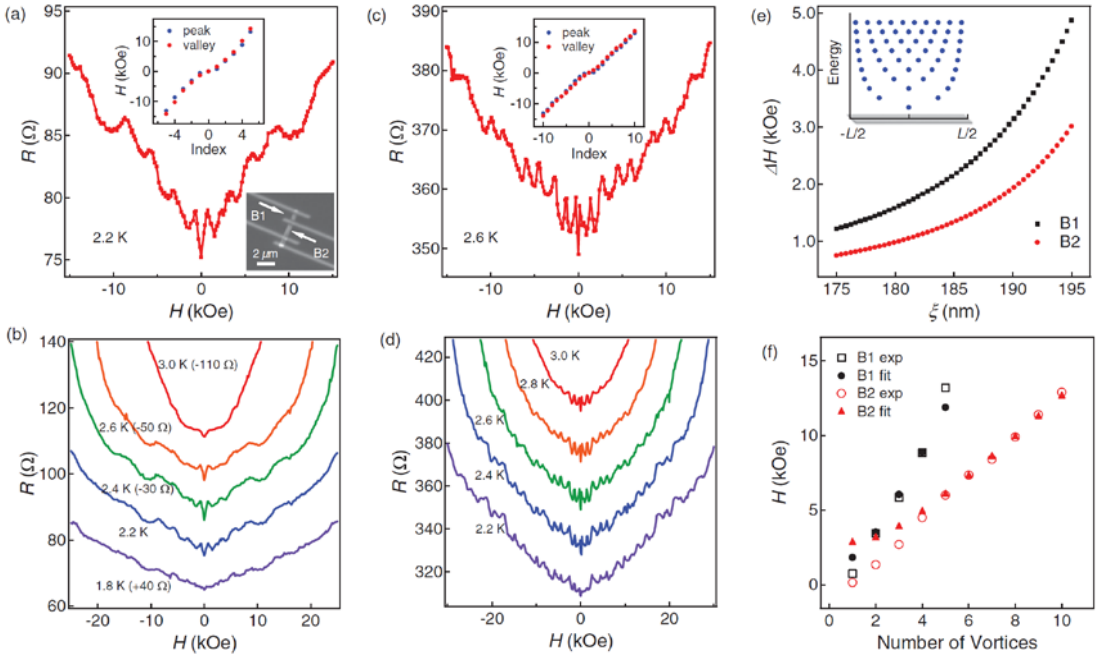

**Fig. 14.** (Color online) (a) MR in device B (channel B1, see lower inset) at T = 2.2 K. The upper inset plots the position of both peaks and valleys. (b) MR in channel B 1 at different temperatures. The data are shifted with respect to the data at T = 2.2 K for clarity. (c) MR in channel B2 at T = 2.6 K. The inset plots the position of both peaks and valleys. (d) MR in channel B2 at different temperatures. The magnetic field in panels (a)–(d) is perpendicular to the nanoribbon plane. (e) Calculated MR oscillation period vs. coherence length (size of vortex) for channels B1 (squares) and B2 (circles). Comparison with the data in Fig. 2(a) and 2(c) yields ξ = 191 nm and 186 nm for channels B1and B2, respectively. The inset is a model calculation showing how vortices distribute in a nanoribbon as the energy (magnetic field) increases. (f) Magnetic field vs. number of vortices. Open symbols are experimental data and filled symbols are fits using the values of ξ obtained in panel (e). [38]

**4.2 Resistance enhancement in SC electrodes**

The proximity effect, while inducing superconductivity in non-superconducting materials adjacent to the SC, simultaneously weakens the superconductivity of the SC near the interface. This would reflect in suppressed transition temperatures and critical fields for the SC as has indeed been observed in the SC-TI film-SC heterostructure we studied. [61] In addition, we also observe an abrupt upturn in the resistance of TI films at the transition temperature/field of the superconducting electrodes.

Our samples are crystalline $Bi_2Se_3$ films grown on sapphire (5 QL) and high-resistivity silicon (200 QL) substrates in ultrahigh-vacuum MBE systems. The high quality has been demonstrated by STM. When the thickness of a TI film decreases, the surface state is thought to be more dominant due to the increasing surface to volume ratio. However, the TI film cannot be too thin as ARPES results showed that when the thickness of $Bi_2Se_3$ is less than 5 QL, the coupling between top and bottom surfaces can open a gap and destroy the surface state.

The two-probe geometry (shown in the inset of Fig. 15(a)) was used to detect the transport properties of $Bi_2Se_3$ films. Bulk indium (In) electrodes and mesoscopic aluminum (Al) and tungsten (W) were used as superconducting electrodes. In dots were directly pressed on the surface of the 5 QL $Bi_2Se_3$ film. The R-T curve is shown in Fig. 15(a). From 300 K to 45K, the curve shows the linear metallic behavior. When the temperature drops below 13.3 K (the resistance minimum), the resistance increases gradually and shows an insulating trend. With further decreasing temperature, an abrupt enhancement at 3.29K (The $T_c$ of bulk In is 3.4K) appears. Figure 15(b) shows the details of the upturn at different low magnetic fields. We can see that the upturn is significantly suppressed with increasing field, and almost disappears when the field is above 200 Oe.

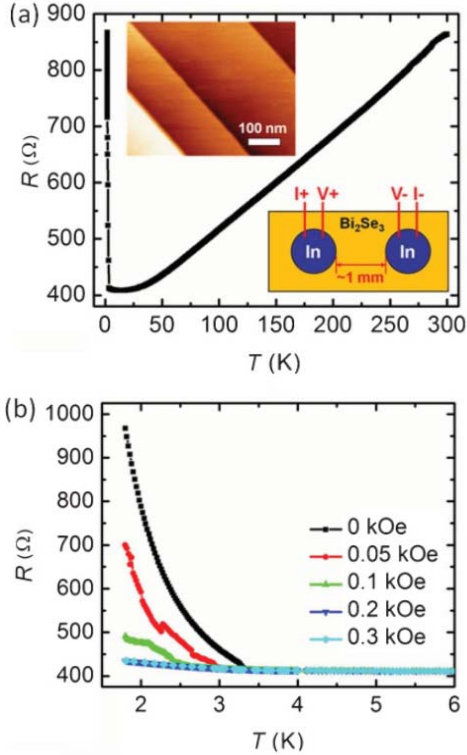

**Fig. 15.** (Color online) The R–T behavior of the 5-nm-thick $Bi_2Se_3$ film contacted by two superconducting In dots. (a) The R vs. T of the 5-QL $Bi_2Se_3$ film from room temperature to low temperature. The left inset is a STM image of the $Bi_2Se_3$ film. The right inset is the measurement structure. (b) The R vs. T at different perpendicular fields. The curves at 0.2 kOe and 0.3 kOe are superimposed. [61]

Figure 16(a) shows the R-H relationship of the 5-QL $Bi_2Se_3$ film when the magnetic field was applied perpendicular to the film. At 4 K, above 2.6 T, MR shows linear behavior, which probably originates from the linear energy-momentum dispersion relationship. [32] But at 1.8K, a MR peak appears near the zero field. Figure 16(b) shows R-H curves obtained at low field and different temperatures. Above the $T_c$ of In, MR is positive, but below $T_c$, MR shows a peak around zero field. Moreover, below $T_c$, even above $H_c$ of In, between 0.2 kOe and 9 kOe, the resistance decreases unexpectedly with the field, which is in contrast to the MR behavior of pure $Bi_2Se_3$ films. The mechanism of this exotic negative MR is still unclear. It might be from the proximity-induced superconductivity in the interface between In electrodes and TI film. Figure 16(c) shows the result of the scan of field from negative to positive, and vice versa. The hysteresis behavior implicates possible ferromagnetic response, but no magnetic contamination was induced in the preparation of samples.

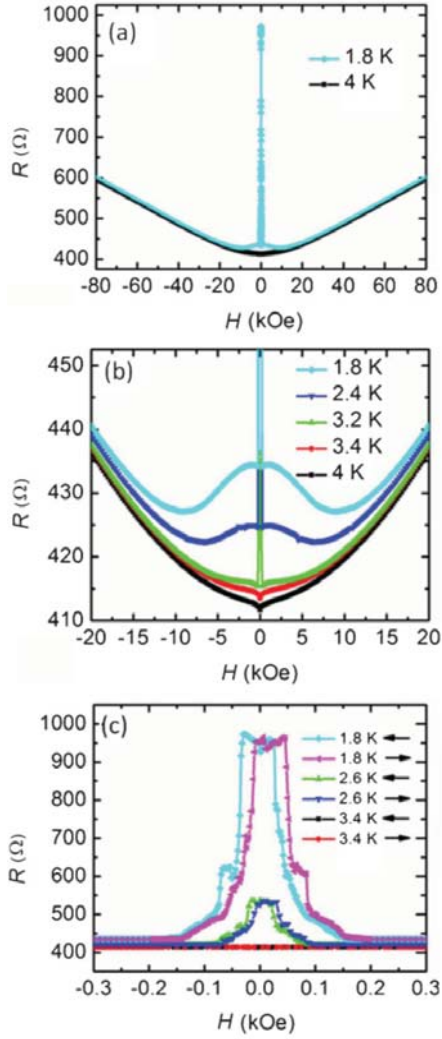

**Fig. 16.** (Color online) The R–H scans of the 5-nm-thick $Bi_2Se_3$ film contacted by In electrodes. (a) Resistance as a function of perpendicular magnetic field at 4 and 1.8 K. (b) Magnified MR for several temperatures. (c) MR peaks near the zero field show terrace structure and hysteresis for scans made below $T_c$. The→arrow indicates that the scan was made from a negative to positive. [61]

Besides bulk indium electrodes, mesoscopic W and Al electrodes were also used for additional experiments. The $Bi_2Se_3$ films in these experiments were 200 QL thick and the distance between two superconducting electrodes was 1 μm. Similar resistance upturn and MR peak around the zero field were observed in these samples (Fig. 17 and Fig. 18). The critical temperature at which the resistance shows an upturn ($T_c$ of the electrode) was noted and compared to the $T_c$ of the superconducting electrodes deposited on an insulator substrate with no TI. It was found that the $T_c$ with the TI is consistently lower than the $T_c$ on an insulating substrate. For example, the upturn of W-TI-W starts around 3.5 K (Fig. 18), which is smaller than $T_c$ of W strips

(4 K-5 K). The $H_c$ of the electrodes is similarly suppressed. For instance, at 2.2 K, the MR peak is totally suppressed at 10 kOe, but the $H_c$ of the W strips is around 80 kOe.

The substantial decreases in $T_c$ and $H_c$ of superconducting electrodes in SC-TI-SC hybrid structures are unexpected as the resistance upturn. To understand this abnormal behavior, we refer to the thoroughly studied semiconductor-SC interfaces in which enhancement in resistance at the transition temperature was observed and explained by Bloder-Tinkham-Klapwijk (BTK) model. [61] In the BTK model, the transparency of the interface is a key factor determining the temperature and bias dependence of the transport. In our case, although the transparencies of our samples are different, the phenomena revealed are essentially the same. The relative magnitude of resistance upturn however, is larger in more transparent TI-SC contact situation (W) compared to the opaque contact (In) which is contrary to the expectations from the BTK model. [62] Since our observations cannot be explained by the traditional BTK model, we propose that the observed phenomenon is connected to the spin-helical surface states of the $Bi_2Se_3$ film or more precisely, is a consequence of the entanglement of bulk and surface transport. Our proposed qualitative explanation is as follows.

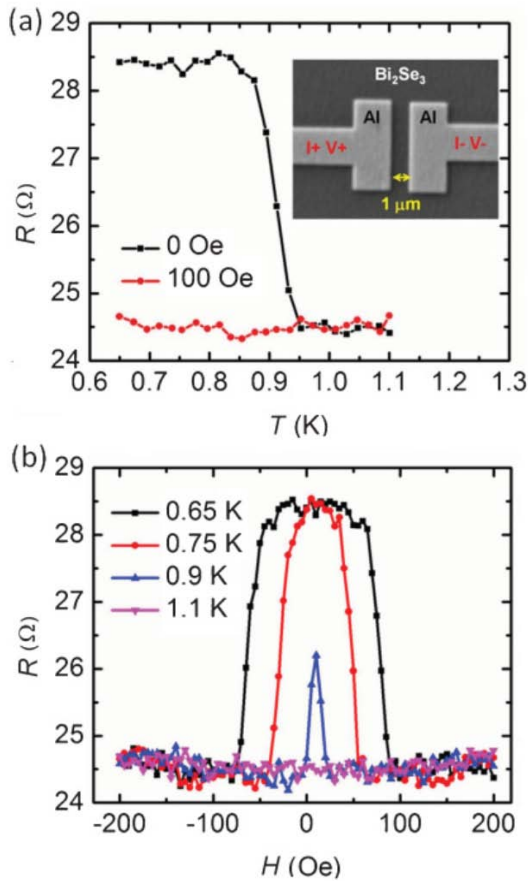

**Fig. 17.** (Color online) Transport behaviors of 200-nm-thick $Bi_2Se_3$ films contacted by superconducting Al electrodes. (a) A sharp resistance enhancement is seen at 0.95 K, which saturates below 0.85 K. An applied magnetic field of 100 Oe suppresses the enhancement. The inset is a SEM image of the Al contacts on the surface of the $Bi_2Se_3$ film. (b) The details of the MR in a small field. The resistance peak at 0.65 K is suppressed under a field of less than 100 Oe, which is much smaller than the $H_c$ (800 Oe) of a 50-nm-thick Al film not contacting with $Bi_2Se_3$ film. The $T_c$ of such an "isolated" Al film is 1.4 K. [61]

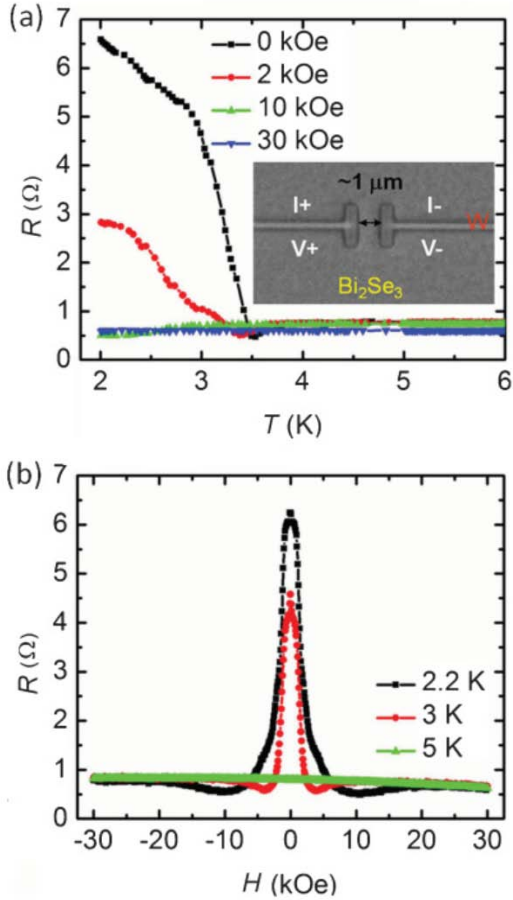

**Fig. 18.** (Color online) Transport behaviors of a 200-nm-thick $Bi_2Se_3$ film contacted by superconducting W electrodes. (a) The R vs. T scans under different magnetic fields. The inset is a SEM image of the W contacts on the surface of the $Bi_2Se_3$ film. (b) MR at different temperatures. When the W electrodes become superconducting, the MR shows a large peak around the zero field. This behavior disappears when the temperature is larger than the $T_c$. The resistance peak is suppressed under a field of ∼10 kOe at 2.2 K, much smaller than the $H_c$ of the W strips (80 kOe). [61]

The TI surface state has a spin-helical structure, which means that the spin and momentum of the electrons are locked. The surface electrons traveling in a particular direction therefore have their spins polarized. When these electrons seek to enter a SC, they need to form Cooper pairs which are composed of two electrons with opposite spins. At the interface of SC and TI, therefore, some of the electrons need to flip their spins. This mechanism may give rise to the resistance upturn and suppress the superconductivity of mesoscopic electrodes. But, for the $Bi_2Se_3$ nanoribbons mentioned in last section, we did not observe the resistance upturn. Two reasons may explain the contradiction. One is the quality of TI films is better than that of nanoribbons and therefore stronger spin-orbit coupling exists in TI films. The second is that the measurement geometry of nanoribbons differs from the geometry of film structures, which may lead to different outcomes.

## 5  Conclusion and outlook

TIs provide a promising new platform for new discoveries uncovering novel physics and also for practical applications. TIs were first predicted by theory and soon thereafter realized by experiments. These novel materials are a new topological state of matter and give us a more profound understanding of condensed matter physics. Although great triumphs, both theoretical and experimental in this area have been achieved, the study of TIs is still in its infancy. On one hand, more and more important discoveries are just emerging in this field. For example, very recently for the first time the Quantum Anomalous Hall (QAH) effect was observed in thin films of chromium-doped $(Bi,Sb)_2Te_3$.[63] This realization of the QAH effect, accompanied by the dissipationless edge states, does not require any magnetic field, so it may lead to low-power consumption electronics. On the other hand, growing TIs with the Fermi level in the gap leading to a real insulating bulk state is a major challenge. Obtaining such TIs is essential not only to the fundamental research but also makes the application of TIs more practical. Different experimental methods have been developed to grow TIs, like MBE,[64,65] chemical vapor deposition (CVD), hybrid physical-chemical vapor deposition (HPCVD),[66] and vapor−liquid−solid (VLS) mechanism.[67] At the same time, the TI family is still open for new members, like semiconductors made by heavy elements and Heusler compounds.[68] All in all, the TI is a rapidly evolving area full of potential. In addition to the intriguing physics described in this review, TIs have many potential applications. For instance, Bi itself has proved to be a good thermoelectric material, so Bi compounds that are TIs may lead to engineering high Seebeck coefficient thermoelectric materials. The combination of TI and SC may generate Majorana fermions and help in fault tolerant topological quantum computing, which is a great challenge in our world today. Some more specific properties have already found technological applications. The linear magnetoresistance of TI can be put to use as magneto sensor, [69] and $Bi_2Se_3$ nanosheet is believed to form good transparent electrodes. [70] Further studying these fascinating materials will reveal novel and exciting physics and may even contribute to reshaping our world.

**Acknowledgement**

We would like to thank those who have contributed to the project over the years, including Moses. H. W. Chan, Qi-Kun Xue, Ke He, Cui-Zu Chang, Nitin Samarth, Jainendra Jain, Mao-Hai Xie, Mingliang Tian, Duming Zhang, Ashley M. DaSilva, Handong Li, Xu-Cun Ma, and Joon Sue Lee – a necessarily incomplete list.